# Anisotropic two-dimensional screening at the surface of black phosphorus

Brian Kiraly[1], Elze J. Knol[1], Klara Volckaert[2], Deepnarayan Biswas[2], Alexander N. Rudenko[3,1,4], Danil A. Prishchenko[4], Vladimir G. Mazurenko[4], Mikhail I. Katsnelson[1,4], Philip Hofmann[2], Daniel Wegner[1], Alexander A. Khajetoorians[1]

1.  *Institute for Molecules and Materials, Radboud University, Nijmegen 6525AJ, The Netherlands*
2.  *Department of Physics and Astronomy, Interdisciplinary Nanoscience Center, Aarhus University, 8000 Aarhus C, Denmark*
3.  *School of Physics and Technology, Wuhan University, Wuhan 430072, China*
4.  *Theoretical Physics and Applied Mathematics Department, Ural Federal University, 620002 Ekaterinburg, Russia*

**Screening in reduced dimensions has strong consequences on the electronic properties in van der Waals semiconductors, impacting the quasiparticle band gap and exciton binding energy. Screening in these materials is typically treated isotropically, yet black phosphorus exhibits in-plane electronic anisotropy seen in its effective mass, carrier mobility, excitonic wavefunctions, and plasmonic dispersion. Here, we use the adsorption of individual potassium atoms on the surface of black phosphorus to vary the near-surface doping over a wide range, while simultaneously probing the dielectric screening via the ordering of the adsorbed atoms. Using scanning tunneling microscopy, we visualize the role of strongly anisotropic screening which leads to the formation of potassium chains with a well-defined orientation and spacing. We quantify the mean interaction potential utilizing statistical methods and find that the dimensionality and anisotropy of the screening is consistent with the presence of a band-bending induced confinement potential near the surface. We corroborate the observed behavior with coverage-dependent studies of the electronic structure with angle-resolved photoemission.**

The role of screening in semiconducting materials with reduced dimensionality has recently seen renewed interest due to the new possibilities arising from isolated single-layer van der Waals (vdW) materials, which can be stacked into complex heterostructures[1-3] and/or tuned through their dielectric environment[4,5]. Given the fact that nearly all van der Waals materials have a hexagonal structure, the resultant screening is highly isotropic in the plane of an individual layer. In contrast, vdW materials with anisotropic screening may host many interesting possibilities for tuning interactions, leading to a pronounced influence on the behavior of collective modes[6,7], the prominence of charged impurity scattering[8], and even the emergence of superconductivity[9] or magnetism[10]. Black phosphorus (BP) is a vdW material that offers the unique opportunity to experimentally explore the effect of anisotropic screening[6,11-14]. The structure and electronic structure of an individual layer BP is highly anisotropic, leading to strong consequences in the charge carrier transport[15,16], particle-hole excitations[17], and plasmonic dispersion[6]. Although internal screening in the long-wavelength limit ($k \rightarrow 0$) is predicted to be isotropic in-plane[14], optical experiments have shown that the excitonic wavefunction in thin BP samples is anisotropic[13]. More sophisticated analytical models and numerical calculations[6] suggest anisotropy in the screened response of few-layer BP, but have not yet been experimentally tested. Despite its fundamental significance for many basic material properties, little is experimentally known about microscopic electrostatic screening in BP.

Here, we quantify the electrostatic screening from the surface layers of BP by looking at its effects on the long-range ordering of charged potassium adatoms. Utilizing scanning tunneling microscopy (STM), we explore the spatial distributions of potassium adatoms as a function of temperature ($T$) and potassium density ($n_K$). STM imaging reveals the development of 1D potassium structures, strongly favoring an orientation along the armchair direction, driven by anisotropic charge screening from the surface layers. Using these images, we quantify the mean interaction potential ($E_m$)[18-20] and corroborate changes to $E_m$ with measurements of the two-dimensional (2D) Fermi contour derived from angle-resolved photoemission spectroscopy (ARPES). We observe very long-ranged, anisotropic screening behavior. While Friedel oscillation-related screening behavior accounts for the long-range ordering in both directions, additional anisotropic short-range correlations are observed along the armchair direction.

An STM image of a clean BP surface after cleaving *in situ* is shown in Fig. 1a. Characteristic intrinsic defects are observed at and below the surface with STM[25], with an areal density of approximately $n_V$ = 3.1x10$^{10}$ cm$^{-2}$. The inset shows the pristine crystal lattice, with the *x* (zig-zag/[010]) and *y* (armchair/[100]) directions labeled. The surface after low-temperature deposition of K atoms ($n_K$ = 7.0x10$^{11}$ cm$^{-2}$) is shown in Fig. 1b. The adatoms are imaged as isotropic protrusions with a bias-dependent apparent height 200 pm < $\Delta z$ < 300 pm. As shown in Fig. 2, the diffusion of K adatoms at the surface of BP occurs between *T* = 4.4 and 5 K, indicating an extremely low lateral diffusion barrier (see also Fig. S1). In order to ensure negligible tip-induced atomic motion, most of the topographic measurements in this work were performed with small currents ($I_t$ < 6 pA).

Adsorption of alkali adatoms usually leads to charge transfer to the substrate. In the case of adsorption on a semiconductor surface, this can result in substantial band bending. This was confirmed here by observing a shift of the band edges with respect to $E_F$ (see Fig. S2 and S3). A sufficiently strong band bending can lead to confinement of the conduction band (CB) states near the surface, as already reported for both BP[21,22], as well as other narrow band gap semiconductors[23,24]. In the case of BP, the resulting two-dimensional electron gas (2DEG) reflects the strong electronic anisotropy of the CB[21] and we can thus expect this electronic structure to be a model system for studying the effect of screening in a strongly anisotropic 2D system.

To study the resultant screened interaction between the positively charged K adatoms, we characterized the relative positions of large numbers (ca. 10,000) of adatoms (*g*(x,y)) in order to calculate the vector-resolved pair distribution function (*g*(x,y)/ *g*$_{ran}$(x,y)), where *g*$_{ran}$(x,y) is the distribution expected for random adsorption sites (see Fig. S4). The mean interaction potential is straightforwardly calculated from the pair distribution function via the reversible work theorem as $E_m(x,y) = -k_B T ln(\frac{g(x,y)}{g_{ran}(x,y)})$. This analysis has previously been used to study dopant distributions in semiconductors[25], the role of dimensionality in screened Coulomb in a variety of materials[20,26], and free carrier-mediated interactions at metallic surfaces[18,19]. In order to probe the effect of screening, we first examined the adsorbate distribution upon adsorption at $T$ = 4.4 K; we then monitored changes to the distribution after annealing the sample to higher temperatures ($T_{anneal}$) in order to overcome surface diffusion barriers (Fig. S1). During the annealing process, the screened interaction potential plays a strong role in the redistribution of K atoms, which can then be observed after cooling again to the measurement temperature of 4.4 K. We also note that the temperature used to derive $E_m$ is a lower bound as we do not probe higher temperatures, although we observe qualitatively similar distributions at higher, uncalibrated, annealing temperatures.

The distribution of K adatoms ($n_K$ = 2.0x10$^{12}$ cm$^{-2}$) after low-temperature deposition ($T$ = 4.4 K) is shown in Fig. 2a. While the large-scale STM image clearly shows a lack of long-range order, the corresponding plot of $E_m$ in Fig. 2e demonstrates short-range repulsion (red) at atomic separations less than 2 nm. This is ascribed to a short-range Coulomb interaction mediated by the substrate. As the sample is annealed, clear anisotropic ordering emerges (Fig. 2b-d). Examining the STM image from Fig. 2b ($T_{anneal}$ = 5.6 K), chain-like structures composed of a few adatoms develop, oriented along the $x$-direction. After annealing to higher temperatures ($T_{anneal}$ = 8.0 K for Fig. 2c and $T_{anneal}$ = 14.8 K for Fig. 2d) the chains extend in length (along $x$-direction) and show a characteristic interchain separation along the $y$-direction. It is noteworthy that the 1D structures orient orthogonal to the direction along which the diffusion barrier is lower, demonstrating that this ordering is not driven by diffusion energy barriers. The development of long-range order seen prominently in Fig. 2c,d is clearly reflected in the plots of $E_m$ (Fig. 2g,h), where deep attractive regions (blue) appear in the $x$-direction and oscillatory interactions emerge along $y$.

To further analyze the potential landscape, line profiles along the $y$-direction ($x$=0) taken from the plots of $E_m$ in Fig. 2e-h are shown in Fig. 2k. The annealing temperature-dependent line profiles along $y$ reveal clear oscillations to distances nearly 40 nm from the origin. While the oscillation amplitude increases with annealing temperature (notably from $T_{anneal}$ = 8.0 K to $T_{anneal}$ = 14.8 K), the oscillatory wavelength remains unchanged. The fact that the ordering is periodic points to an interaction potential that is strongly influenced by Friedel-oscillations along $y$, as seen for isotropic interactions in other systems[18,19]. As we show later, the observed periodicity is indeed consistent with Fermi wavelength along the $y$-direction ($\lambda_{F,y}$). Also, doping-induced changes of the ordering can be explained by corresponding modifications of the Fermi surface as a function of electron filling.

The behavior of $E_m$ along $x$ (Fig. 2j), is strikingly different from $y$ (Fig. 2k). The interatomic spacing between atoms in the 1D chains can be ascribed to a prominent potential minimum approximately 4 nm from the origin, as well as a second weaker minimum at 8 nm. However, these structures cannot be reconciled with Friedel oscillations, because they appear at distances much shorter than $\lambda_{F,x}$ (see Fig. 2j). Explaining the atomic separation along the chain thus requires a more sophisticated treatment of the K-doped BP dielectric function to describe the screened Coulomb interactions, along with a consideration of the preferred adsorption positions on the lattice. However, weak manifestations of Friedel-type interactions can still be observed in the $x$ direction, because half the Fermi wavelength ($\lambda_{F,x}/2 = 15\pm6$ nm) determined from ARPES coincides with the characteristic finite length of the chains (see Fig. S5).

The interpretation of the observed ordering in terms of Friedel oscillations can be further confirmed by coverage-dependent measurements in which the filling and Fermi vectors of the 2DEG are changed. Fig. 3 shows a series of interaction potentials obtained from different potassium coverages, collected after annealing at temperatures between 14 and 18 K. At $n_K = 7.0 \times 10^{11}$ cm$^{-2}$, the mean interaction potential (Fig. 3a – for STM images, see Fig. S6) clearly lacks the pronounced anisotropy seen for higher coverages in Fig. 3b-d and Fig 2h. The potential is repulsive at short range and shows a weakly attractive region at around 11 nm. We note that at this coverage we cannot deduce the potential with high precision. With increasing coverage ($n_K \geq 1.4 \times 10^{12}$ cm$^{-2}$), there is a qualitative change which can be explained by the CB minimum crossing $E_F$. Beyond this coverage, $E_m$ remains qualitatively similar, but as we show in more detail below, small changes in the characteristic periodicity appear that can be linked to the doping-dependent filling of the 2DEG (inset Fig. 4a).

For a detailed analysis of the doping-dependent changes, as well as the dimensionality of the screening, we analyze the decay of the oscillation in the y-direction (Fig. 4a). A radially symmetric interaction potential derived from an electron gas with an isotropic Fermi surface can be described by[27]:

$$E_m(R) = -C \frac{\cos(2k_F R + \delta)}{(k_F R)^m}$$  eq. 1

where $C$ is a constant, $k_F$ is the characteristic Fermi surface wave vector, $R$ is the distance to the scattering center, $\delta$ is the carrier phase shift upon scattering, and $m$ is the spatial dimensionality. We fit the oscillatory $E_m$ data in Fig. 4a to eq. 1, obtaining the scattering wave vectors $q_y = 2*k_{F,y}$, phase shifts, and dimensionality $m$. For the highest doping level ($n_K = 1.8 \times 10^{13}$ cm$^{-2}$, see Fig. S7), the best fit is obtained using $m = 2$, resulting in $q_y = 0.28 \pm 0.02$ Å$^{-1}$. The profiles taken from $n_K = 2.0 \times 10^{12}$ cm$^{-2}$ (red points) and $n_K = 1.4 \times 10^{12}$ cm$^{-2}$ (blue points) result in $q_y = 0.08 \pm 0.01$ Å$^{-1}$ and $q_y = 0.07 \pm 0.01$ Å$^{-1}$, respectively. As seen in Fig. 4a, this data shows much slower decay in the oscillatory $E_m$. In fact, taking $m = 1$ provides a better fit to this data (red and blue lines, respectively), indicating a quasi-one-dimensional screening along $k_y$. This observed scattering dimensionality could be related to anisotropic charge puddling around the K dopants or Fermi surface nesting[28], but the origin is not clear from this formalism.

In order to independently confirm both the Fermi contour dimensions and dimensionality of the states near $E_F$, we perform doping-dependent ARPES measurements (see Fig. S9). The band structure of K-doped BP ($n_{2D} = 1.0 \times 10^{14}$ cm$^{-2}$) is shown in Fig. 4b,c. To confirm the 2D confinement of the bands near $E_F$, data is collected as a function of photon energy, demonstrating a clear lack of $k_z$ dispersion (Fig. S10). The ARPES measurements reaffirm the indications from STM measurements that K-doped BP hosts a 2D electron system at the surface, corroborating the previous conclusions from ARPES experiments[21,22]. Furthermore, the 2D bands show anisotropy and effective masses roughly consistent with expectations for both monolayer and bulk BP[29]. Using Luttinger's theorem to extract the carrier density ($n_{2D}$) from ARPES band structure, we compare the observed Fermi surface vectors ($k_{F,x}$ and $k_{F,y}$) with those obtained from STM (under the assumption $n_K = n_{2D}$), as shown in Fig. 4d. The agreement between the experimental methods at both low and high $n_K$ is excellent. Furthermore, the band anisotropy seen at larger $n_K$ (orange points) matches both the ARPES observations and theoretical predictions well[30,31].

In conclusion, we demonstrate that K doping of the BP surface reveals the anisotropic in-plane screening and quasi-1D ordering of K adatoms mediated by the quasi-2D electron system induced by band-bending. By examining large arrays of interacting K adatoms on the surface of BP, we show that the anisotropic screened Coulomb response governs the formation of 1D potassium chains and that the near-surface confined charge carriers mediate extremely long-ranged interchain interactions (> 40 nm along the *y*-direction), at relatively low carrier densities ($n_K$ = 2.0x10$^{12}$ cm$^{-2}$). The anisotropic response of the system persists to high carrier densities ($n_K$ = 1.8x10$^{13}$ cm$^{-2}$), allowing us to correlate the STM results with photoemission experiments. We note that the unusual interatomic spacing in a given K chain cannot be accounted for considering screening effects at long wavelength and requires a more sophisticated picture of the K-doped BP dielectric function. The results show that conduction electron-mediated interactions in BP are strongly anisotropic and extremely long-ranged, with significant ramifications for the electron dynamics and excitonic excitations in doped BP and phosphorene.

**Acknowledgements**

The authors would like to acknowledge scientific discussions with Malte Rösner. E.K. and A.A.K. also acknowledge the VIDI project: 'Manipulating the interplay between superconductivity and chiral magnetism at the single atom level' with project number 680-47-534 which is financed by NWO. B.K. and A.A.K. acknowledge support from the European Union's Horizon 2020 research and innovation programme under grant agreement No. 751437. A.A.K acknowledges support from the European Research Council (ERC) under the European Union's Horizon 2020 research and innovation programme (grant agreement No 818399, SPINAPSE). M.I.K. acknowledges support from the JTC-FLAGERA project GRANSPORT. This work was supported by VILLUM FONDEN via the Centre of Excellence for Dirac Materials (Grant No. 11744). V.G.M. acknowledges support from the Ministry of Science and Higher Education of the Russian Federation, Project No. 3.7372.2017/8.9.

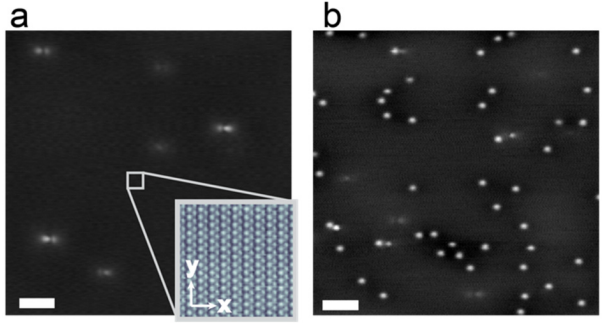

FIG. 1. Clean and K-doped black phosphorus. (a) STM constant-current image showing pristine black phosphorus with characteristic vacancies ($V_s$ = -400 mV, $I_t$ = 20 pA, scale bar = 10 nm). The inset shows an atomically resolved image of the black phosphorus surface ($V_s$ = -100 mV, $I_t$ = 40 pA). The notation used here is armchair = $x$ = [100] and zig-zag = $y$ = [010]. (b) K-doped black phosphorus at areal adatom density $n_K$ = 7.0*10$^{11}$ cm$^{-2}$ after low-temperature deposition at $T$ = 4.4 K. K adatoms appear as isotropic protrusions with apparent height of approximately 200 pm ($V_s$ = 1 V, $I_t$ = 10 pA, scale bar = 10 nm).

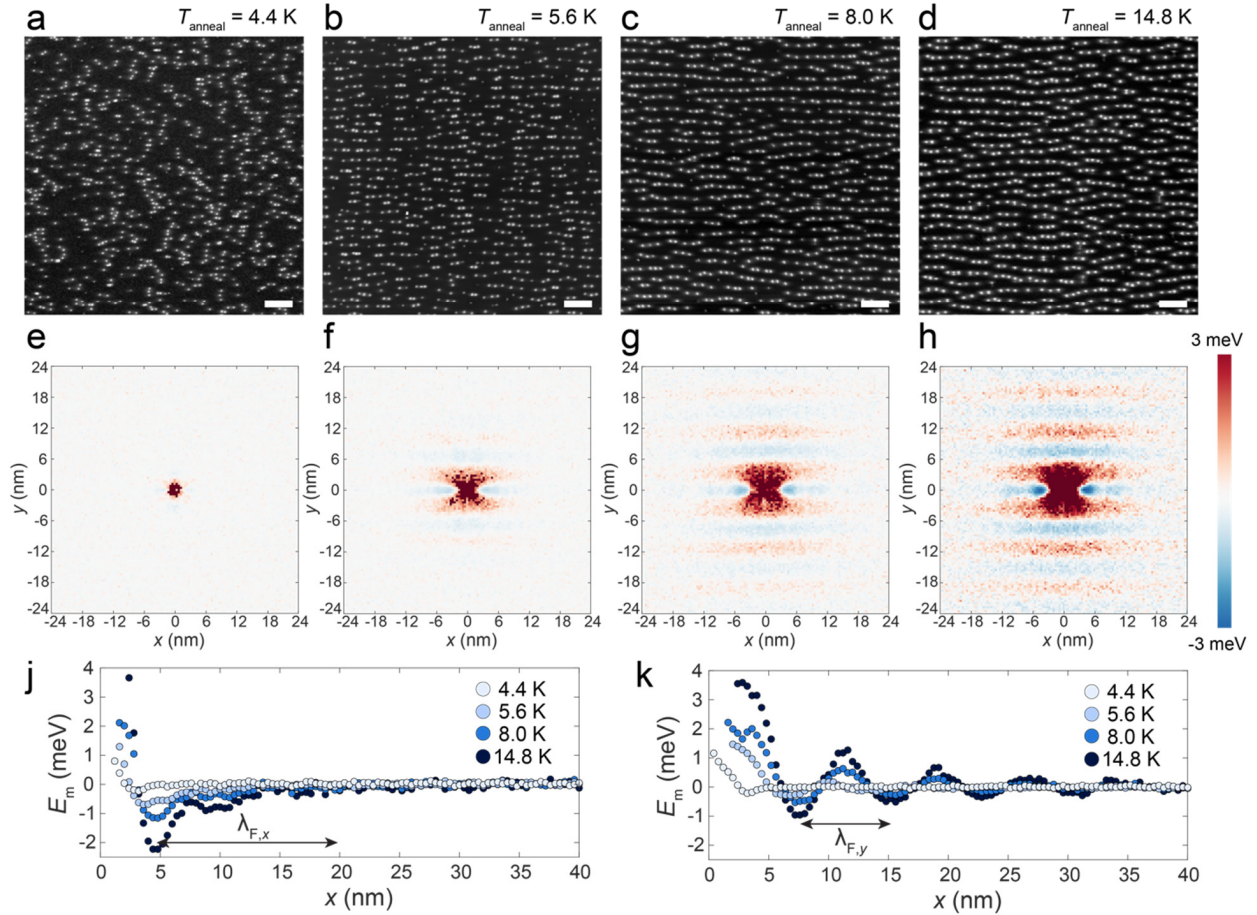

FIG. 2. Temperature-dependent distribution of K adatoms. (a) Large-scale STM constant-current image of K-doped BP (adatom density $n_K$ = 2.0x10$^{12}$ cm$^{-2}$) after low-temperature deposition at $T$ = 4.4 K ($V_s$ = -1 V, $I_t$ = 3 pA, scale bar = 20 nm). (b-d) STM constant current images taken at $T$ = 4.4 K after annealing the K-doped BP sample for ten minutes to $T_{anneal}$ = 5.7, 8.0, and 14.8 K, respectively ($V_s$ = -1 V, $I_t$ = 3 pA, scale bar = 20 nm). (e-h) Mean interaction potentials for single K dopant at (0,0) calculated from images (a-d), respectively. Line cut from mean interaction potential (j) along $x$ and (k) along $y$.

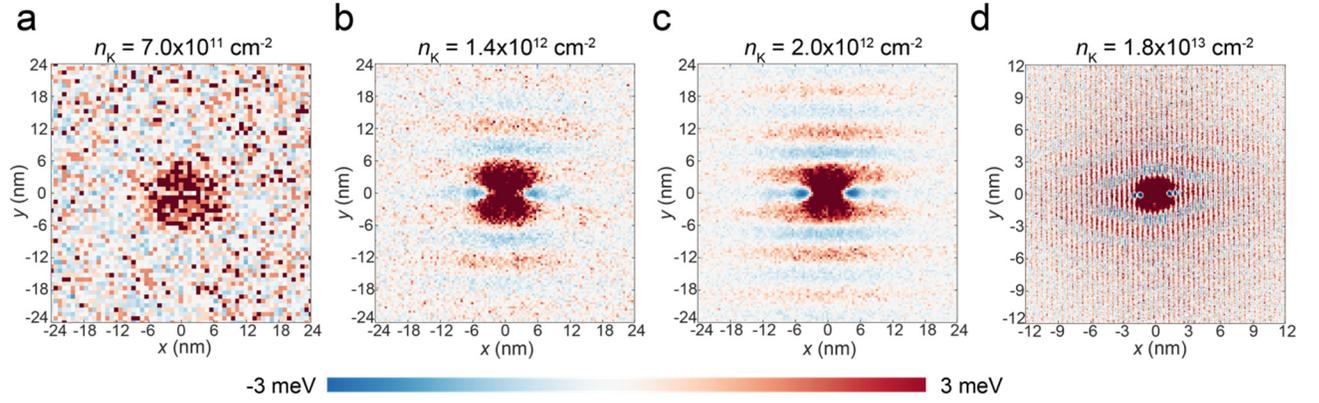

FIG. 3. Density-dependent mean interaction potentials. Mean interaction potentials for varying K adatom density ($n_K$) are shown after annealing the samples to 13.4 K (lowest density), 14.8 K (medium and high density), and approximately 18.0 K (highest density). (a) Mean interaction potential for lowest K density ($n_K = 7.0 \times 10^{11}$ cm$^{-2}$) showing isotropic screening behavior. (b) Mean interaction potential at $n_K = 1.4 \times 10^{12}$ cm$^{-2}$, revealing the onset of screening anisotropy. (c), (d) Mean interaction potential for $n_K = 2.0 \times 10^{12}$ cm$^{-2}$ and $n_K = 1.8 \times 10^{13}$ cm$^{-2}$, respectively.

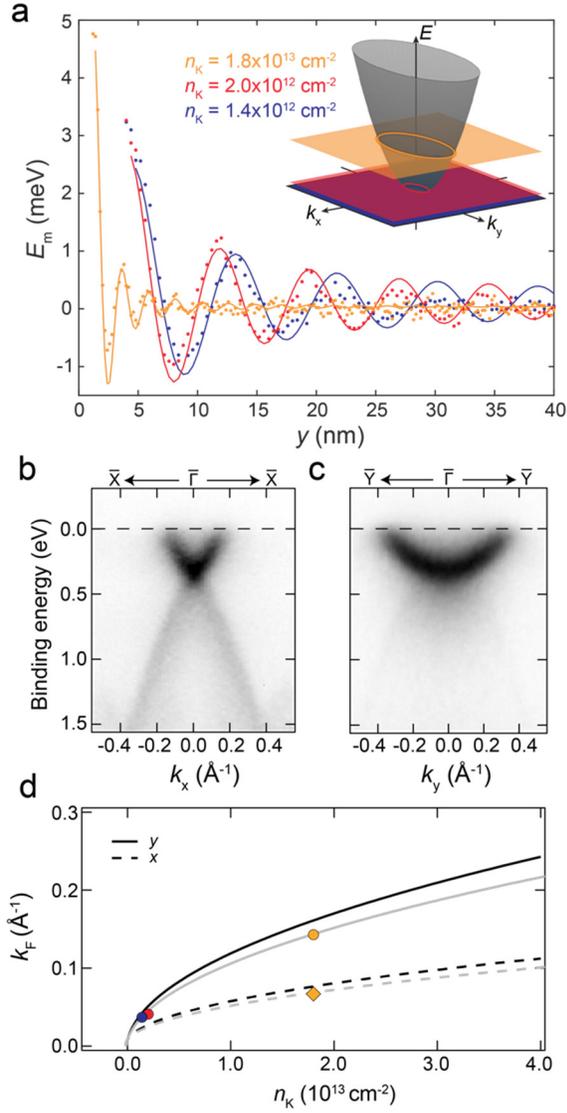

FIG. 4. Relation between electron filling and screening. (a) Line profiles of the oscillatory mean interaction potential along $y$ for samples doped with $n_K = 1.4 \times 10^{12}$ cm$^{-2}$ (dark blue), $n_K = 2.0 \times 10^{12}$ cm$^{-2}$ (red), and $n_K = 1.8 \times 10^{13}$ cm$^{-2}$ (orange). (Inset) Illustration of the anisotropic CB of BP with colored slices indicating doping levels from Figs. 3 and 4a. (b) and (c) ARPES data taken after dosing K onto BP showing a 2D electron-like band along (c) $\bar{\Gamma}$-$\bar{X}$ and (d) $\bar{\Gamma}$-$\bar{Y}$ directions. (d) Comparison of the Fermi wave vectors along $\bar{\Gamma}$-$\bar{X}$ and $\bar{\Gamma}$-$\bar{Y}$ as a function of potassium density ($n_K$) for ARPES data (solid and dashed lines) and STM data (points). Dopant density for ARPES data was calculated using Luttinger's theorem under the assumption that a single electron per K atom is transferred to the BP surface layer (black) and 0.8 electrons are transferred (gray).

Supplementary information

# Anisotropic two-dimensional screening at the surface of black phosphorus

Brian Kiraly[1], Elze J. Knol[1], Klara Volckaert[2], Deepnarayan Biswas[2], Alexander N. Rudenko[3,1,4], Danil A. Prishchenko[4], Vladimir G. Mazurenko[4], Mikhail I. Katsnelson[1,4], Philip Hofmann[2], Daniel Wegner[1], Alexander A. Khajetoorians[1]

1. Institute for Molecules and Materials, Radboud University, Nijmegen 6525AJ, The Netherlands
2. Department of Physics and Astronomy, Interdisciplinary Nanoscience Center, Aarhus University, 8000 Aarhus C, Denmark
3. School of Physics and Technology, Wuhan University, Wuhan 430072, China
4. Theoretical Physics and Applied Mathematics Department, Ural Federal University, 620002 Ekaterinburg, Russia

## Table of Contents

## Methods

*Scanning Tunneling Microscopy/Spectroscopy*

STM/STS measurements were performed under ultrahigh vacuum (< 1x10$^{-10}$ mbar) conditions with an Omicron low-temperature STM at a base temperature of 4.4 K, with the bias applied to the sample. All STM images in the text and supplementary information were acquired by means of constant-current feedback. Electrochemically etched W tips were used for measurements; each tip was treated *in situ* by electron bombardment, field emission, as well as dipped and characterized on a clean Au(111) surface. STS was measured using a lock-in technique to directly measure d$I$/d$V$; a modulation of $V_{mod}$ = 2-6 mV at a frequency of $f_{mod}$ = 4.23 kHz was applied to the bias signal. Black phosphorus crystals were purchased from HQ graphene and subsequently stored in vacuum (< 1x10$^{-8}$ mbar). The crystals were cleaved under ultrahigh vacuum conditions at pressures below 2x10$^{-10}$ mbar, and immediately transferred to the microscope for *in-situ* characterization. Potassium was evaporated from a SAES getter source directly onto the cold sample inside the STM with $T_{STM}$ = 4.4 K for the entire duration of the dosing procedure. Annealing was accomplished by shining white light into the STM for a period of 10 minutes, while monitoring the temperature directly on the STM stage. After annealing, the samples were again scanned at $T_{STM}$ = 4.4 K.

*Angle-Resolved Photoemission Spectroscopy*

The ARPES experiments were performed at the SGM3 beamline of the synchrotron radiation source ASTRID2[1] at Aarhus, Denmark. The sample was kept at a temperature of 30 K and measured at a photon energy of 104 eV with a total energy and *k* resolution of 40 meV and 0.01 Å$^{-1}$, respectively. Potassium was deposited on black phosphorous by evaporation from a SAES getter source.

The Fermi wave vectors were determined from the peak positions of the momentum distributions curves at the Fermi energy. The curves displayed in Fig 4(e) represent a 4th order polynomial fit to these peak positions. The carrier concentration was calculated using Luttinger's theorem[2], by assuming an elliptical Fermi surface with $k_F$ values deduced from the MDC fit.

## Diffusion of Potassium on Black Phosphorus

The diffusion barrier of K on undoped BP was determined from first-principles calculations for diffusion along the [100] (armchair or *x*-direction) and [010] (zig-zag or *y*-direction) directions[3] (Supplementary Fig. 1). The diffusion barrier was estimated by calculating the potential energy landscape and finding the minimum-energy transition path along the *x*- and *y*-direction. To this end, we considered a $(3a_x \times 4a_y)$ supercell containing a single K atom adsorbed on a BP monolayer, and used density functional theory (DFT) to calculate the total energies for different positions of the K atom. DFT calculations were performed using the projected augmented-wave method (PAW)[4] as implemented in the Vienna *ab initio* simulation package (VASP)[5,6]. Exchange and correlation effects were taken into account within the gradient approximation (GGA) in the parametrization of Perdew-Burke-Ernzerhof (PBE)[7]. An energy cutoff of 300 eV for the plane-wave basis and the convergence threshold of $10^{-6}$ eV were used in the self-consistent solution of the Kohn-Sham equations, which checked to be sufficient to obtain numerical accuracy. Pseudopotentials were taken to include 3*s* and 3*p* valence electrons for P atoms, as well as 3*s*, 3*p*, and 2*s* valence electrons for the K atom. Vertical separation between the layers was set to 30 Å. The Brillouin zone was sampled in two-dimensions by a uniform distribution of **k**-points on an (8 × 8) mesh. The calculated barrier along the *y*-direction is approximately 37 meV, roughly similar to potassium on graphene[8-10]. The barrier along the *x*-direction is nearly an order of magnitude larger (287 meV). These differences indicate that the diffusion of K adatoms occurs almost exclusively along the *y*-direction at the temperatures reached in these experiments (*T* < 20 K). We note that the diffusion barriers can also show doping dependence, and therefore primarily serve as approximations for the diffusion energies.

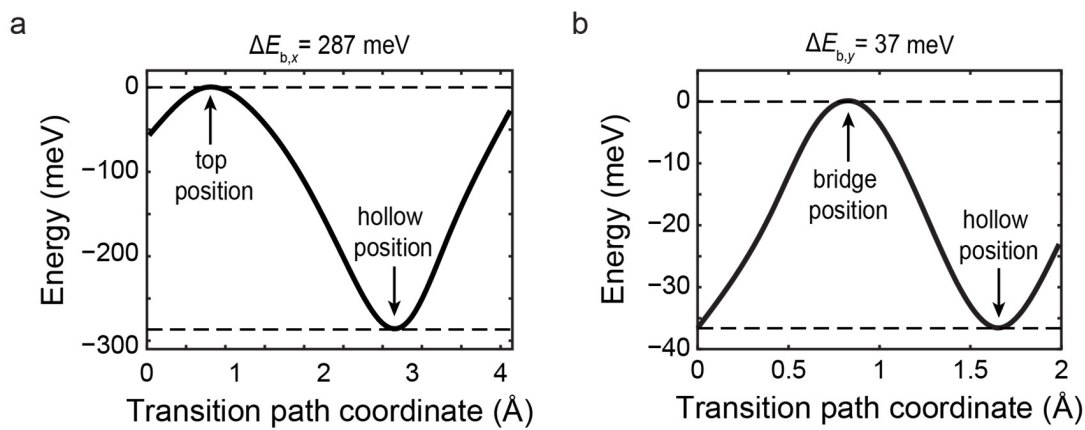

**Supplementary Figure 1. Diffusion of K on BP surface.** Calculated diffusion barriers of K adatoms along the (a) x-direction (armchair) and (b) y-direction (zig-zag).

Potassium Doping of Black Phosphorus

To study the effect of K doping on the band structure of BP, scanning tunneling spectroscopy (STS) and density functional theory (DFT) calculations were performed. The results of both methods show that deposition of K atoms on BP leads to *n*-doping.

STS spectra taken on pristine and K-doped BP ($n_K$ = 2.4x10$^{12}$ cm$^{-2}$) reveal a rigid shift of the band gap of approximately 0.3 V to lower energy upon K deposition, consistent with *n*-doping (Supplementary Fig. 2). The positions of the valence band (VB) maxima and conduction band (CB) minima were determined with the same method as in earlier work on pristine BP[11]. For pristine BP (Supplementary Fig. 2a) that analysis results in a band gap of $E_g$ = 0.33 ± 0.01 eV, with the VB maximum around $E_{VB}$ = -0.045 eV and the CB minimum near $E_{CB}$ = 0.286 eV, which is consistent with previous work[11]. For K-doped BP (Supplementary Fig. 2b) the measured band gap is $E_g$ = 0.33 ± 0.02 eV, with the VB maximum approximately at $E_{VB}$ = -0.30 eV and the CB minimum near $E_{CB}$ = 0.03 eV. There is no significant change in the size of the band gap, even though both the VB minimum and CB maximum of the band gap shift down in energy approximately 0.3 eV. The spectra measured on the doped sample were taken at least 2 nm away from any K adatoms. Due to the low diffusion barrier of the K adatoms, it was very difficult to take spectra close to the K adatoms without tip-induced motion. Therefore, we are unable to take spectra across the entire surface and derive an average band gap value to precisely determine the band onset. This possibly explains why many $E_{CB}$ measurements are slightly above $E_F$. Furthermore, we note that visible differences in apparent height near the potassium atoms suggest charge puddling indicative of changes to the local density of states.

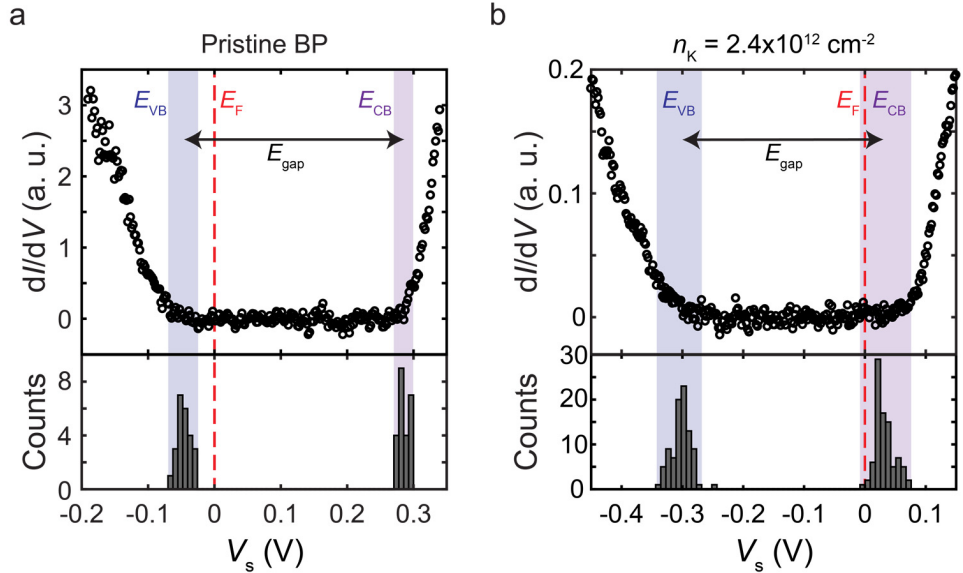

**Supplementary Figure 2. Scanning tunneling spectroscopy of pristine and K-doped BP.** (a) Top: d$I$/d$V$ spectrum on pristine BP ($V_{mod}$ = 4 mV). Bottom: histogram of the extracted VB maxima and CB minima from 24 spectra. Shaded regions indicate the measured range of the VB maxima (blue) and CB minima (purple). (b) Top: d$I$/d$V$ spectrum on K-doped BP ($V_{mod}$ = 4 mV, $n_K$ = 2.4x10$^{12}$ cm$^{-2}$). Bottom: histogram of the extracted valence band maxima and conduction band minima from 89 spectra, taken at least 2 nm away from any K adatoms. Shaded regions again indicate the measured range of the VB maxima (blue) and CB minima (purple).

The band structure was calculated using DFT for two different concentrations of K atoms. For this purpose, we used supercells with dimensions ($a_x \times 2a_y$) and ($3a_x \times 4a_y$), which correspond to the carrier concentrations $n_K$ = 3.5 × 10$^{14}$ cm$^{-2}$ and $n_K$ = 5.7 × 10$^{13}$ cm$^{-2}$, respectively. Other parameters of the DFT calculations were taken the same as for the calculations of the diffusion barriers. The projection of the electronic bands on the states of the K atom was done using the formalism of maximally localized Wannier functions[12] implemented in the wannier90 package[13]. The calculated band structures are shown in Supplementary Fig. 3. For both K adatom densities the valence bands are shifted ~0.25 eV below the Fermi level ($E_F$). Such a band shift is consistent with the VB shift observed in STS (Supplementary Fig. 2b). The primary difference observed with dopant density is the energy of the electron-like band at Γ, which shifts from ~0.25 eV below $E_F$ at $n_K$ = 5.7x10$^{13}$ cm$^{-2}$ to ~0.75 eV below $E_F$ at $n_K$ = 3.5x10$^{14}$ cm$^{-2}$. As shown in the color scale, this electron-like band is comprised mainly of phosphorus $p_z$ states, with ~30% contribution from K $s$ states. We note that the anisotropy in the band structure is similar for 2D and 3D BP, although this calculation does not include the band-bending potential. This band has similar anisotropy to the BP CB, and is responsible for the charge carrier screening seen with the STM.

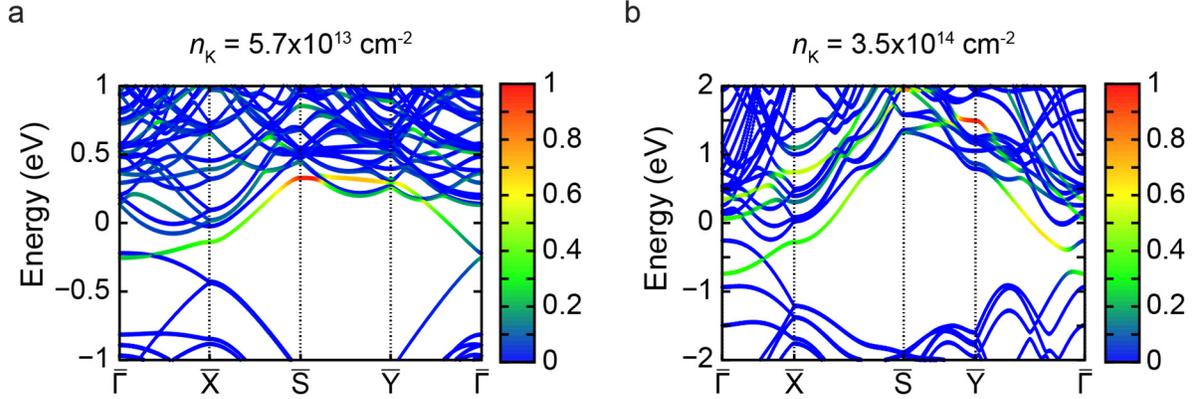

**Supplementary Figure 3. *Ab-initio* calculations for K-doped BP.** Band structure of doped BP for carrier concentrations of (a) $n_K = 5.7\times10^{13}$ cm$^{-2}$ and (b) $n_K = 3.5\times10^{14}$ cm$^{-2}$. The color scale denotes the contribution of the different orbitals: blue bands stem solely from P orbitals, while red indicates contribution from K orbitals.

Extracting the Mean Interaction Potential

To extract the mean interaction potential ($E_m$), which describes electrostatic interactions between adatoms, a statistical analysis was performed using STM constant-current images according to previous literature[14-26]. The first step of the analysis is to determine the position $r_i(x_i, y_i)$ of the center of each K atom in an image. Then, the relative positions of all other K adatoms (summing over all $j$ atoms $r_j(x_j, y_j)$) with respect to an individual K adatom ($r_i$) is determined: $\vec{r}_{ij} = (x_i - x_j, \; y_i - y_j)$. The resulting collection of vectors, after repeating the procedure for each adatom, can be plotted as a 2D histogram, as shown in Supplementary Figure 4b. This is the distribution function $g(x, y)$, counting the number of K adatoms at a given distance $\vec{r}$ from any other K adatom. To extract $E_m$ from the distribution function (Supplementary Figure 4c), we apply the reversible work theorem[24]:

$$E_m(x,y) = -k_\text{B}T \ln\left(\frac{g(x,y)}{g_{ran}(x,y)}\right) \qquad (1)$$

Here, $k_\text{B}$ is the Boltzmann constant, $T$ is the temperature at thermodynamic equilibrium and $g_{ran}(x,y)$ is the distribution function for a random selection of adatoms[24]. Approximately 10,000 adatoms were measured in each independent determination of $E_m$ within the main manuscript.

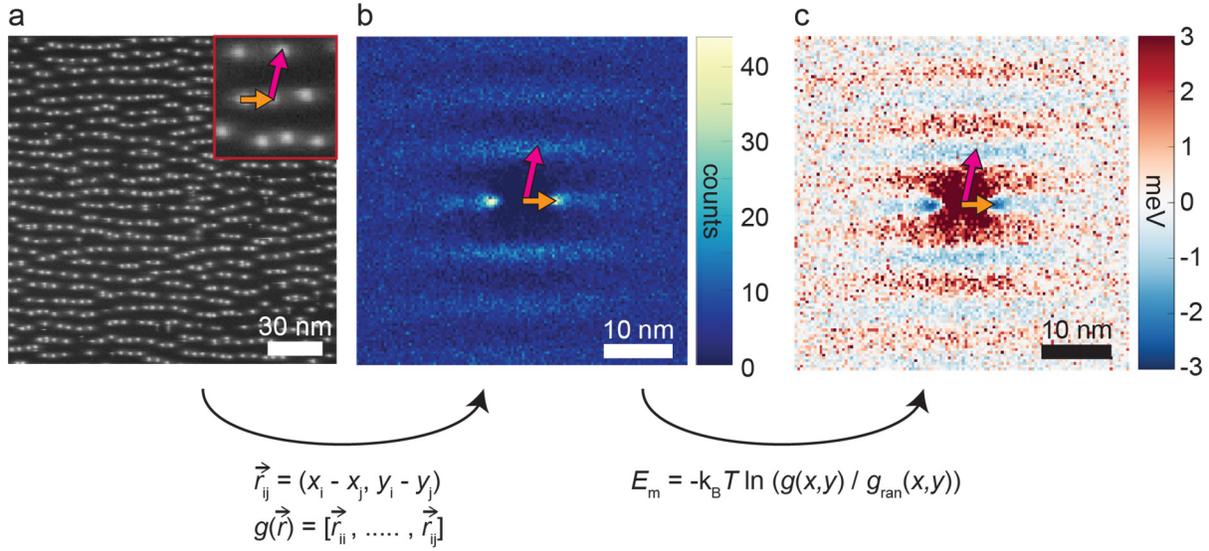

**Supplementary Figure 4. Calculation of the mean interaction potential.** (a) STM constant-current image of K on BP. In the inset, two vectors $r_{ij}$ are given between two K adatoms. (b) Vector-resolved histogram of all vectors $r_{ij}$. The two arrows correspond to the two vectors shown in the inset in (a). (c) Mean interaction potential from the full image shown in (a), containing roughly 3,000 atoms. The mean interaction potentials in the main text are generated with roughly 10,000 atoms.

Interchain Periodicities

To understand the characteristic length scales governing the length and separations between atomic K chains in Fig. 2d, we binarized the data from Fig. 2d using a simple threshold technique (Supplementary Fig. 5a). This was done to effectively isolate the chains from the background. To evaluate the characteristic length scales, we performed a fast-Fourier transform (FFT) of the data in Supplementary Fig. 5a, shown in Supplementary Fig. 5b. The characteristic interchain spacing along $y$ is clearly evident in the peak of the FFT (Supplementary Fig. 5d) at $k_y$ = 0.080 Å$^{-1}$. While less pronounced, there is also a small peak in the FFT along $k_x$ at $k_x$ = 0.034 Å$^{-1}$. Both peaks nearly coincide with the dimensions of the Fermi surface ($k_{F,x}$ and $k_{F,y}$), pointing to the fact that the oscillatory Friedel oscillations play an important role in setting the interchain spacing along both $x$ and $y$.

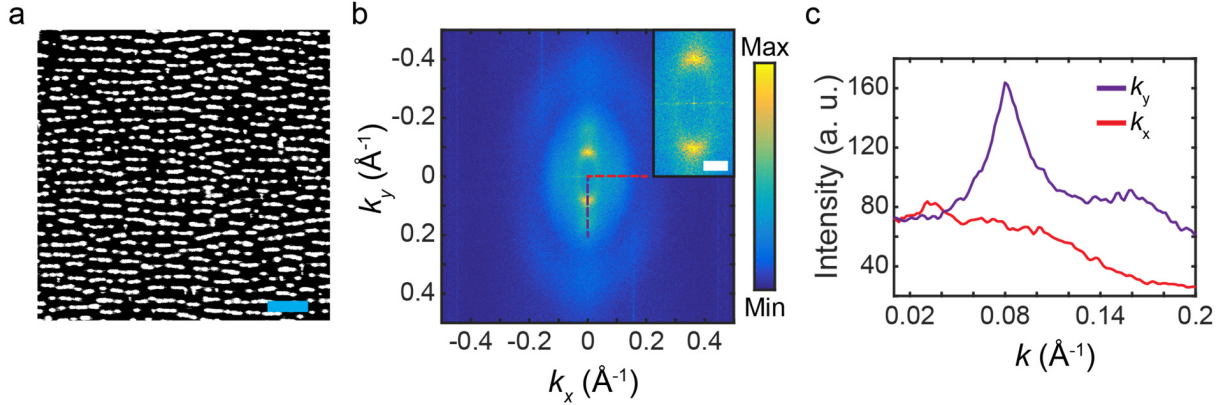

**Supplementary Figure 5. Analysis of interchain periodicities.** (a) Binarized STM topography at $n_K$ = $2.0 \times 10^{12}$ cm$^{-2}$ after annealing at 13.4 K. Scale bar 30 nm, $V_s$ = -1 V, $I_t$ = 3 pA. (b) FFT of the binarized topography in (a). The dashed lines correspond to the line cuts in (c) and the inset shows a zoom of the center of the image. The scalebar in the inset corresponds to 0.04 Å$^{-1}$. (c) Line cuts from the Fourier transform seen in (b) for both the $k_x$ and $k_y$ directions.

Coverage-Dependent Annealing

Constant-current STM images showing the temperature evolution of the K adatom distribution for $n_K$ = $7.0 \times 10^{11}$ cm$^{-2}$, $n_K$ = $1.4 \times 10^{12}$ cm$^{-2}$, and $n_K$ = $2.0 \times 10^{12}$ cm$^{-2}$ are shown in Supplementary Figure 6. The lowest adatom density ($n_K$ = $7.0 \times 10^{11}$ cm$^{-2}$) is shown in Supplementary Fig. 6a-d; the temperature-dependent evolution clearly shows an absence of anisotropy in the adatom distribution, reflected in the plot of $E_m$ shown in Fig. 3a of the main text. At higher adatom densities ($n_K$ = $1.4 \times 10^{12}$ cm$^{-2}$ and $n_K$ = $2.0 \times 10^{12}$ cm$^{-2}$) anisotropic ordering is observed as is shown in Fig. 2 in the main text. Despite small differences, the general characteristics of the distributions are not strongly modified between $n_K$ = $1.4 \times 10^{12}$ cm$^{-2}$ and $n_K$ = $2.0 \times 10^{12}$ cm$^{-2}$, as seen in Supplementary Fig. 6e-h and Supplementary Fig. 6j-m, respectively.

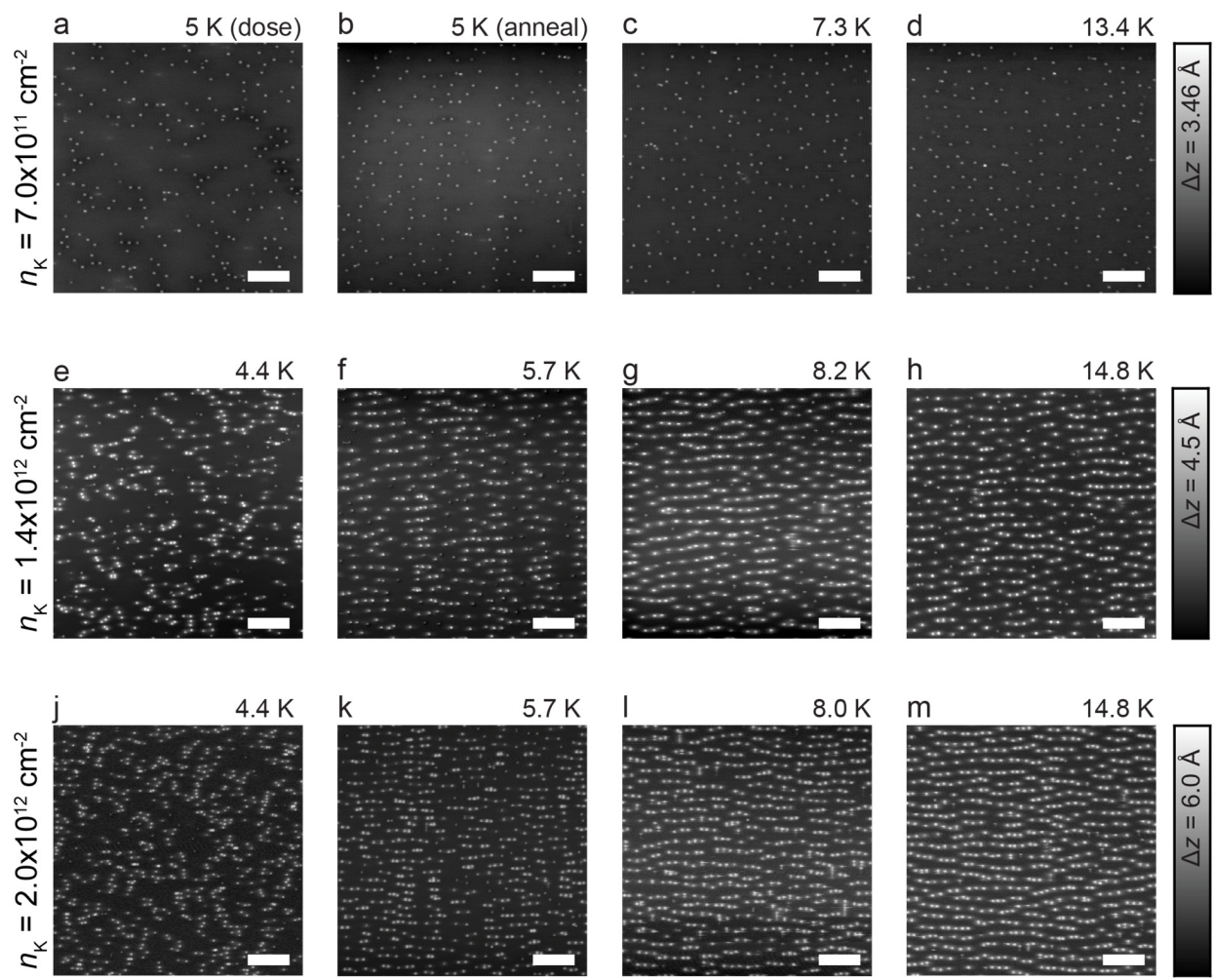

**Supplementary Figure 6. Various coverages of K/BP with annealing.** (a-d) STM topographies with the lowest K density ($n_K = 7.0 \times 10^{11}$ cm$^{-2}$), not annealed (5 K) and after annealing to 5 K, 7.3 K and 13.4 K, respectively. (e-h) STM topographies at $n_K = 1.4 \times 10^{12}$ cm$^{-2}$, not annealed (4.4 K) and after annealing to 5.7 K, 8.2 K and 14.8 K, respectively. (j-m) STM topographies at $n_K = 2.0 \times 10^{12}$ cm$^{-2}$, not annealed (4.4 K) and after annealing to 5.7 K, 8.0 K and 14.8 K, respectively. For all images: scale bar 30 nm, $V_s = -1$ V, $I_t = 3 – 4.5$ pA.

## Highly Doped Black Phosphorus

A constant-current STM image of K-doped BP at $n_K = 1.8 \times 10^{13}$ cm$^{-2}$ is shown in Supplementary Fig. 7 after annealing to $T_{anneal} = 18$ K. This dopant density corresponds to the plot of $E_m$ shown in the main text in Figure 3d. Here, strong variations in the adatom distribution are seen in the dramatically reduced nearest-neighbor separations along the x-direction. These variations are also observed in the plot of $E_m$ shown in both Fig. 3d and, with greater detail, in Supplementary Fig. 8a. Here, deep attractive potentials appear along the x-direction at ~1.3 and 1.7 nm from the origin. These potential wells correspond to separations between the atoms of 3 and 4 zig-zag rows. In fact, the vertical features in Supplementary Fig. 8a result from discretization of the K binding sites, as K adsorption is confined to the hollow sites. In addition to the potential wells, elliptical oscillatory interactions are observed at larger distances (Supplementary Fig. 8a). We attribute these oscillatory interactions to charge carrier screening of the charged K dopants. To confirm the experimental observations, we used a simple model to study the charge redistribution within doped phosphorene around a single point charge at the surface. To calculate the charge density oscillations, we used the following expression:

$$\delta n(\mathbf{r}) = \frac{Ze}{(2\pi)^2} \int d^2q \left(\frac{1}{\epsilon(\mathbf{q})} - 1\right) e^{i\mathbf{q}\cdot\mathbf{r}}, \tag{2}$$

which describes the charge density response of a point charge impurity $n(\mathbf{r}) = Ze\delta(\mathbf{r})$ embedded into a 2D dielectric media, where Z is the impurity charge. $\epsilon(\mathbf{q})$ is the static dielectric function in reciprocal space, which in the random-phase approximation has the form[27]:

$$\epsilon(\mathbf{q}) = 1 + V(q)\Pi(\mathbf{q}), \tag{3}$$

where $V(q) = e^2/2\epsilon_0 q$ is the bare 2D Coulomb interaction in reciprocal space, and $\Pi(\mathbf{q})$ is the 2D polarizability in the static limit. To calculate $\Pi(\mathbf{q})$ for BP, we used a model expression which was proposed by T. Low et al.[28] for a BP monolayer:

$$\Pi(\mathbf{q}) = g_{2D}\Re\left[1 - \sqrt{1 - \frac{8\mu/\hbar^2}{q_x^2/m_x + q_y^2/m_y}}\right], \tag{4}$$

where $g_{2D} = \sqrt{m_x m_y}/\pi\hbar^2$ is the 2D density of states, $\mu$ is the chemical potential, and $m_x = 0.15 m_0$ and $m_y = 1.0 m_0$ are the electron effective masses along $q_x$ and $q_y$ directions, respectively.

The Fourier transform in Eq. (2) was calculated numerically on a uniform (4000 x 4000) $q$-point mesh using a cutoff of $q_{cut} \cong 40\pi/a$, where $a$ is the lattice parameter of BP, being $a_y = 3.313$ Å for the zigzag direction and $a_x = 4.374$ Å for the armchair direction. The resulting charge distribution for a charge carrier density of ($n_{2D}$ = 8.0x10$^{12}$ cm$^{-2}$) is shown in Supplementary Fig. 8b. The oscillatory charge density has both a similar length scale and ellipticity as the experimentally measured interactions. The calculations confirm the experimental picture of charge carrier-mediated anisotropic interactions, although they neglect the effects of the band bending potential and that the charge sources are K adatoms.

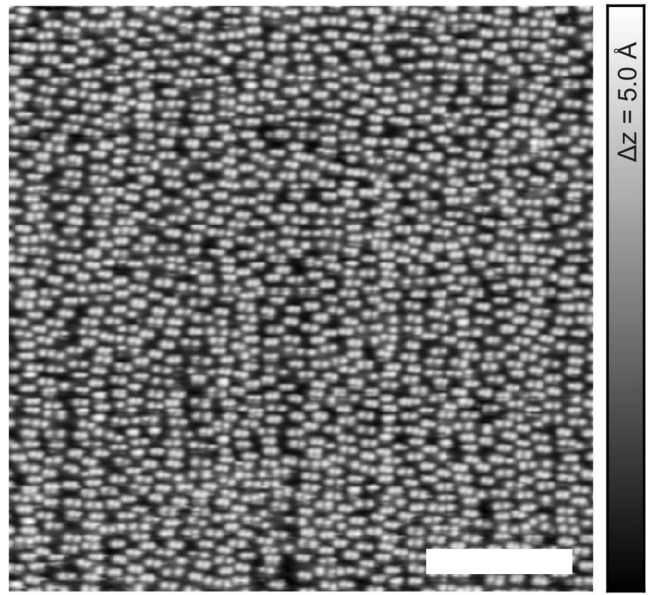

**Supplementary Figure 7. Scanning tunneling microscopy of highly K-doped BP.** Adatom density: $n_K$ = 1.8x10$^{13}$ cm$^{-2}$. Scale bar 30 nm, $V_s$ = -2 V, $I_t$ = 3 pA.

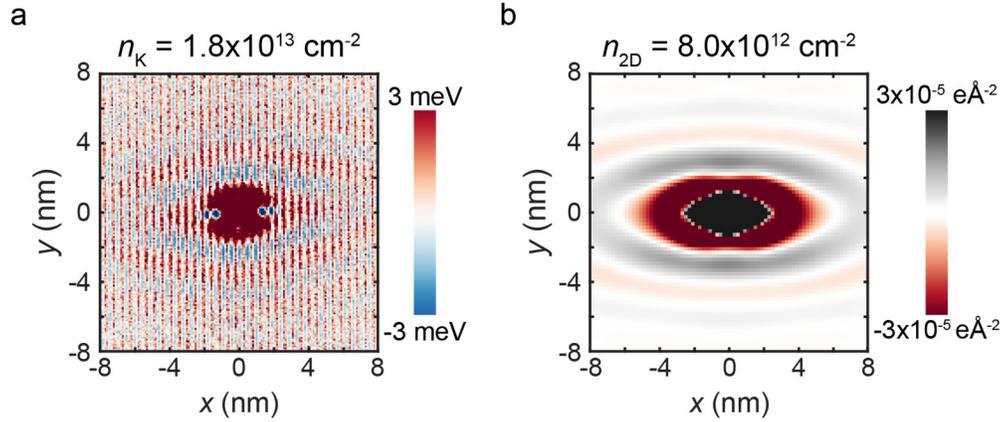

**Supplementary Figure 8. Comparison between theoretical and experimental Friedel oscillations.** (a) Mean interaction potential with $n_K = 1.8 \times 10^{13}$ cm$^{-2}$ showing elliptical oscillatory envelope characteristic of free-carrier scattering on the Fermi surface. (b) Numerical calculation for charge-density redistribution around a single point charge on monolayer BP with $E_F$ located 0.05 eV above the CB minimum (corresponding carrier density $n_{2D} = 8.0 \times 10^{12}$ cm$^{-2}$).

## Band Structure of Doped Black Phosphorus

Angle-resolved photoemission spectroscopy (ARPES) was used to study the dopant density-dependent band structure of BP. The results are qualitatively similar to those from previous experiments[3,29,30]. Supplementary Fig. 9 shows cuts of the band structure around $\bar{\Gamma}$ along $k_x$ and $k_y$ for pristine BP and several stages of potassium doping. Upon doping, multiple changes to the band structure occur: (1) the valence band shifts downward (away from $E_F$), consistent with STS measurements and (2) an electron pocket appears at $E_F$ and shifts downward with increasing doping. As expected, the band curvature of both the VB and CB is strongly anisotropic, characteristic of the differing effective masses along the $x$- and $y$-directions.

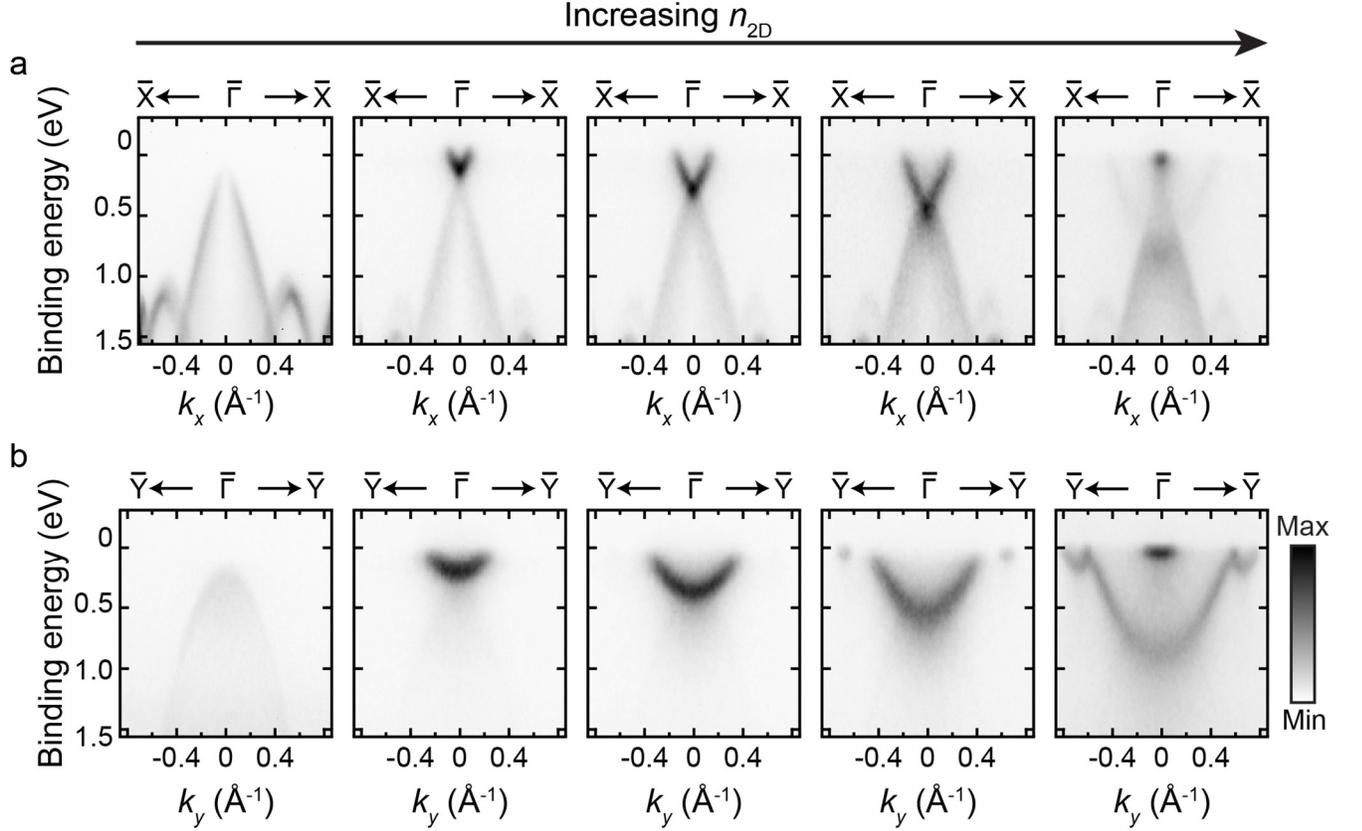

**Supplementary Figure 9. Band structure at varying K coverage.** ARPES spectra on BP along (a) $k_x$ and (b) $k_y$ before deposition, and K-doped BP for $n_{2D}$ = 2.0x10$^{13}$ cm$^{-2}$, $n_{2D}$ = 6.3x10$^{13}$ cm$^{-2}$, $n_{2D}$ = 1.1x10$^{14}$ cm$^{-2}$, and $n_{2D}$ = 4.3x10$^{14}$ cm$^{-2}$.

The band dispersion around the $\bar{\Gamma}$ point is plotted along $k_x$, $k_y$ and $k_z$ in Supplementary Figure 10 for the case of a high dopant density. In Supplementary Figure 10a,b, the CB minimum has shifted below the VB maximum, reaching the so-called surface band inversion seen previously[29-31]. The 2D nature of these bands is confirmed by examining the band dispersion along $k_z$ (Supplementary Figure 10c,d). In bulk BP, the VB (and CB) strongly disperse along the z-direction due to the interlayer interactions. Yet, as seen in Supplementary Figure 10d, the CB observed at $\bar{\Gamma}$ is non-dispersive along $k_z$. The lack of $k_z$ dispersion confirms that a 2D electronic state is present at the Fermi energy in K-doped BP.

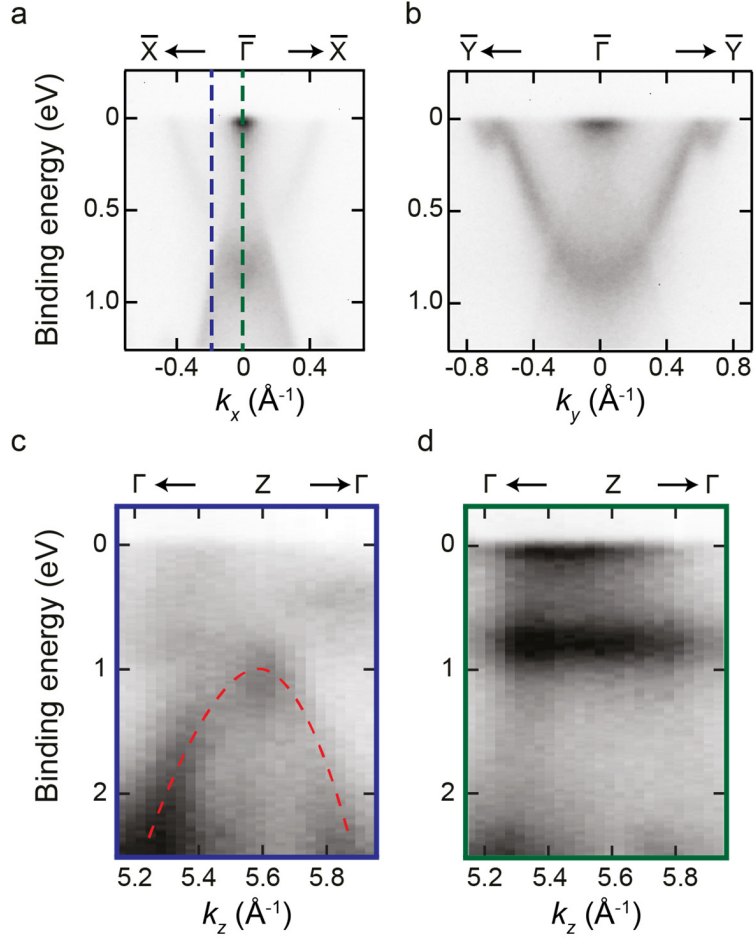

**Supplementary Figure 10. Out-of plane band dispersion for K-doped BP.** ARPES of K-doped BP along (a) $k_x$, (b) $k_y$ and (c-d) $k_z$. The blue and green dashed lines in (a) correspond to the fixed $k_x$ values at which the band structure was measured along $k_z$, shown in (c) and (d) respectively.

We fit the CB seen in ARPES (Supplementary Fig. 11) as a function of K-doping to extract the carrier density-dependent effective mass along $\bar{\Gamma} - \bar{Y}$ (Supplementary Table 1). The table shows a small renormalization (< 15%) of the electron effective mass for high doping. Overall, the strong band anisotropy (Supplementary Fig. 9) and larger effective masses along $\bar{\Gamma} - \bar{Y}$ are consistent with expectations for both BP monolayer and bulk[32-34].

| $n_{2D}$ (x$10^{14}$ cm$^{-2}$) | Effective Mass ($m^*/m_e$) |
|---|---|
| 0.20 | 1.71 ± 0.02 |
| 0.63 | 1.57 ± 0.01 |
| 1.1 | 1.39 ± 0.01 |
| 4.3 | 1.41 ± 0.01 |

**Supplementary Table 1. Carrier density-dependent effective masses.** Values for the effective mass corresponding to the carrier density ($n_{2D}$) listed, where the uncertainty is derived from the fit. We observe a slight decrease of the effective mass of the electron-like band for higher doping levels.

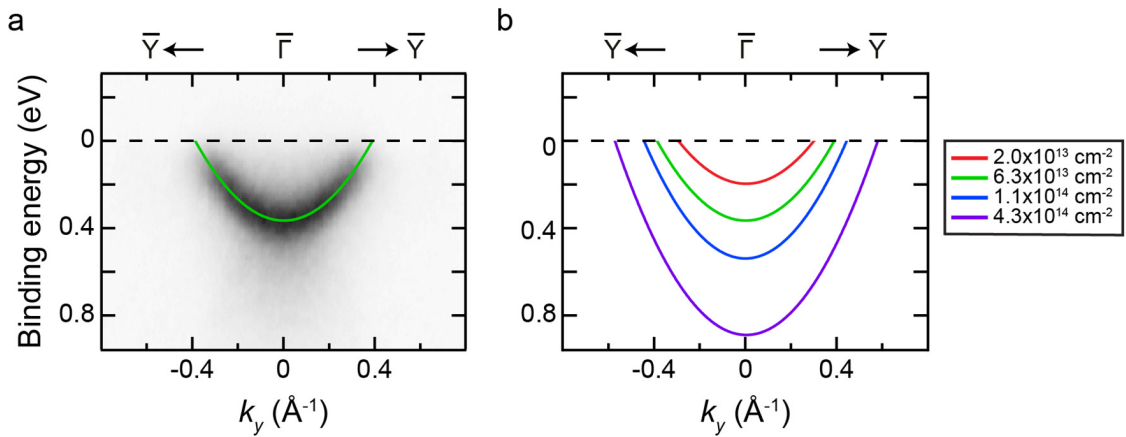

**Supplementary Figure 11. Effective mass of electron-like band upon deposition.** (a) ARPES dispersion of the surface state along $\bar{\Gamma} - \bar{Y}$ at $n_{2D}$ = 6.3x10$^{13}$ cm$^{-2}$. The green curve indicates a parabolic fit to the peak positions of the energy distribution curves. (b) Parabolic fits for different carrier concentrations shown in red ($n_{2D}$ = 2.0x10$^{13}$ cm$^{-2}$), green ($n_{2D}$ = 6.3x10$^{13}$ cm$^{-2}$), blue ($n_{2D}$ = 1.1x10$^{14}$ cm$^{-2}$) and purple ($n_{2D}$ = 4.3x10$^{14}$ cm$^{-2}$).